# Is Science Inevitable?

Large-Scale Data Reveals That Breakthroughs Are Multiple, Not Singular

Linzhuo Li†, Yiling Lin‡, and Lingfei Wu‡

†Dept. of Sociology, Zhejiang University, Zhejiang, 310058, P.R. China
‡School of Computing and Information, The University of Pittsburgh, Pittsburgh, Pittsburgh, PA 15260
*Corresponding author. E-mail: linzhuoli@zju.edu.cn  (L.L) and liw105@pitt.edu (L.W)

For centuries, scientific breakthroughs were thought to be the result of rare genius, with lone figures such as Newton, Darwin, and Einstein credited with singular discoveries[1]. Merton challenged this view, arguing that science advances through multiple, and therefore inevitable, discoveries, driven by cumulative knowledge growth and the intellectual zeitgeist, rather than individual brilliance[2]. Despite its influence, this claim remained largely untested at scale. Here, we use large-scale citation data and a new breakthrough metric to systematically evaluate whether scientific breakthroughs are as inevitable as Merton proposed. Our findings reveal that breakthroughs indeed emerge in multiples rather than as singular events, reinforcing the structural inevitability of scientific progress. This research complements prior work on scientific stagnation, which has focused on the declining innovation of the average paper[3,4], by showing that while breakthroughs are exceptionally rare, they remain inevitable under the right intellectual conditions.

Merton's theory is difficult to test because multiple discoveries are rare, independent events, requiring case-by-case identification. His dataset contained only 264 cases [2], while Simonton's, the largest to date, identified 1,434 multiples—but only after a decade of manual searching[5,6]. The availability of large citation graphs offers hope, as they allow systematic tracking of how a common ancestor paper inspires downstream research, sometimes leading to independent discoveries that "step on each other's toes." However, citation links alone are insufficient for identifying multiples, as papers cite the same reference for different reasons and may belong to unrelated fields. Instead, we need a way to measure *functional equivalence*—cases where independent papers build on the same idea to address similar problems.

The Disruption Index (D-index)*, initially developed to measure technical innovation[7] and later introduced to assess scientific breakthroughs[8], provides a robust measure of functional equivalence in scientific discoveries. Unlike the common belief that it captures absolute innovation[9], our analysis shows that the D-index quantifies a paper's ability to *displace* its most-cited reference within the same research problem, like light bulbs replacing candles—building on past solutions while ultimately making them obsolete. When multiple papers independently displace the same reference, they form a pool of multiple discoveries. We present mathematical analysis, empirical validation, and case studies to demonstrate how the D-index captures functional equivalence in scientific breakthroughs.

The D-index measures how a focal paper (*f*) displaces its most cited reference, as follow:

---

* The datasets for reproducing the main findings of the D-index are available on Harvard Dataverse: https://doi.org/10.7910/DVN/VE3AFX.

$$D_f = \frac{N_i - N_j}{N_i + N_j + N_k} = \frac{N_i - N_j}{N_i + N_j} \frac{1}{1 + \frac{N_k}{N_i + N_j}} \approx d_f \frac{1}{1 + b_f}$$

where $(N_i + N_j) = C_f$ represents the total citation impact of the focal paper, including $N_i$ (papers citing only the focal) and $N_j$ (papers citing both the focal paper and its reference). We define $d_f = (N_i - N_j)/C_f$ as the *Local Displacement Factor,* measuring the normalized disparity between these two types of citing groups. The term $N_k$ represents papers citing only the focal paper's references but not the focal paper itself, approximating the total citations to its references ($C_{total}$). Based on Zipf's law, this can be further approximated as $C_{total} \approx k C_{max}$, where $C_{max}$ is the citation impact of the focal paper's most cited reference. The coefficient $k$ is approximated as $k \approx (b+1)/(a-1)$, where $a$ is the scaling exponent of Zipf's law and $b$ is a fitted constant. Our analysis confirms $k$ is a small constant (~2.5), leading to $b_f \sim C_{max}/C_f$, where $b_f$ is the *Knowledge Burden Factor*—reflecting the ratio of the most cited reference's citation impact to that of the focal paper. Inspired by the "burden of knowledge" theory[10], this suggests that a truly disruptive paper must surpass citation impacts of the "giant's shoulders" it stands upon.

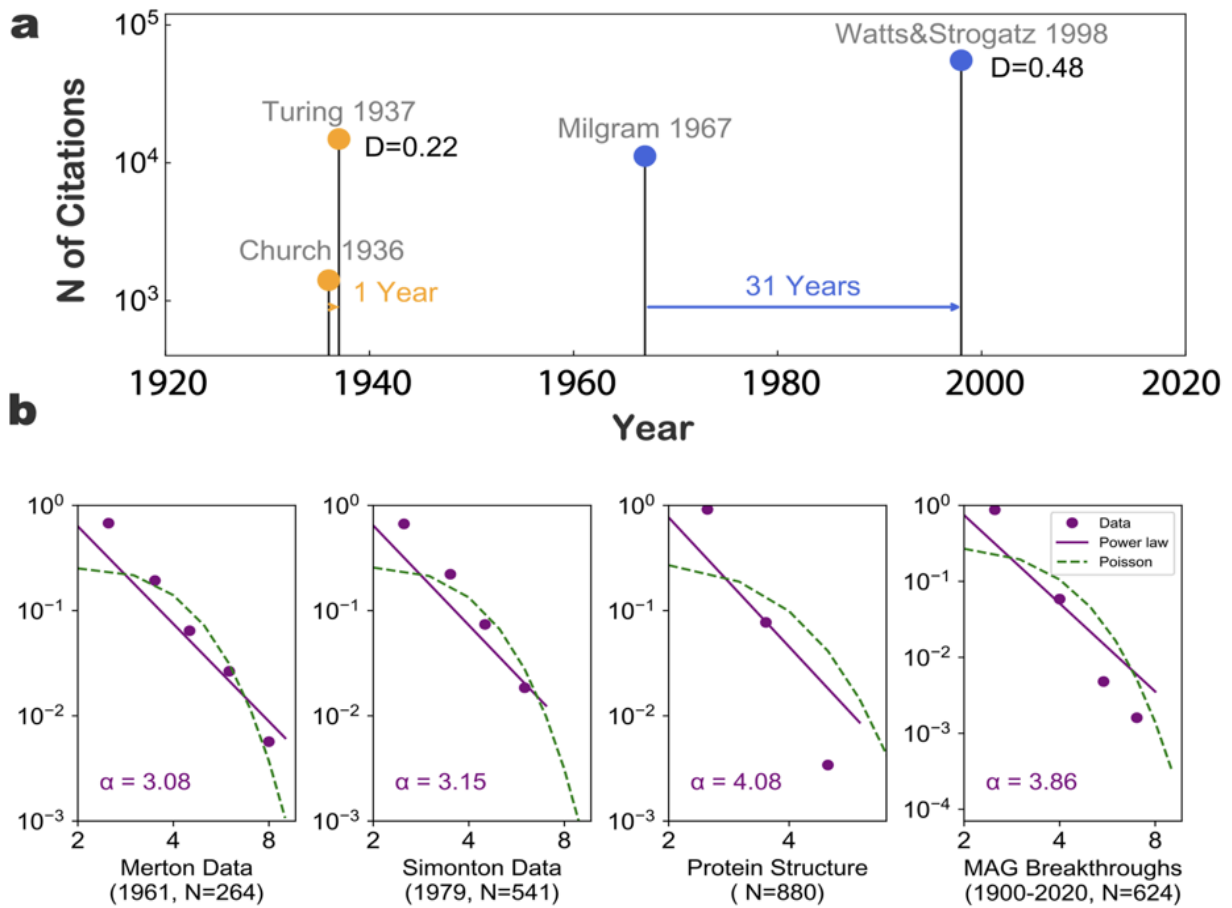

**Figure 1. Multiple Discoveries are not independent and follow power law distributions. a.** Examples show a displacement event (e.g., Church to Turing and Milgram to Watts & Strogatz). **b.** The distribution of multiples, including Merton's (*N*=264, 1600–1950), Simonton's (*N*=1,434, 1350–1990), the Protein Data Bank (*N*=880, 1999–2017), and multiple breakthroughs in MAG (*N*=624, 1900–2020), identified by Llama-3.1-405B-Instruct. The purple line shows a power law fit, while the green line shows the Poisson fit.

Next, we examine whether high D-index papers displace their most-cited reference within the same research problem. Analyzing over 40 million journal articles (1900–2020) from the archived Microsoft Academic Graph (now part of OpenAlex), we selected the top 2% of papers by D-index (D>0.21). Each paper was assigned field labels (e.g., Discrete Mathematics or Molecular Biology) from MAG's 292-category taxonomy, with an average of two unique labels per paper. If displacement were random, the probability of a paper and its most-cited reference sharing the same field would be 0.015. However, empirical analysis shows a 0.62 probability—over 40 times higher than expected. This strong alignment suggests that breakthroughs typically emerge through focused advancements within the same intellectual lineage rather than random recombinations of prior knowledge[11].

To gain deeper insights, we re-analyzed data from a 2019 global breakthrough survey[8], which collected 190 breakthrough papers nominated by 20 scientists across nine disciplines and five countries, with an average D-index of 0.21—notably higher than the 0.1 average for 877 Nobel Prize-winning papers (1902–2009)[8]. Our review of the top 10 nominated breakthroughs and their most-cited references reveals a consistent pattern of displacement within the same topic. For example, in biology, Watson & Crick (1953) displaced Pauling & Corey (1953) with the correct double-helix model of DNA, replacing the incorrect triple-helix, and in computer science, Turing (1937) displaced Gödel (1931) by reformulating incompleteness theorems through the Turing machine model. Combined with mathematical and empirical analysis, these cases confirm the functional equivalence between high-D papers and their most-cited reference, and thus among high-D papers that share the same most-cited reference.

After establishing the D-index as a proxy for multiple discoveries, we tested Merton's multiple discovery model by selecting 12,564 high-impact (≥100 citations) and disruptive (D>0.2) papers (1900-2020) along with their most-cited references, identifying sibling breakthroughs sharing the same reference. Analyzing the distribution of multiple disruptive discoveries, we found a long-tail pattern, contradicting Merton's Poisson model[2,6]. To rule out D-index bias, we examined three additional datasets: Merton's historical records (N=264, 1600-1950), Simonton's dataset (N=1,434, 1350-1990), and the Protein Data Bank (N=880, 1999–2017), which identified multiples using protein similarity scores (>50%)[12]. All exhibited the same long-tail distribution. While this challenges Merton's mathematical framework, it supports his core claim that scientific discoveries are inevitable as they arise from historical context rather than isolated genius.

As our case study reveals, in 1953, Franklin & Gosling rejected Pauling & Corey to independently propose the double-helix DNA model[13], and in 1936, Church reformulated Gödel's incompleteness theorems using λ-calculus[14]. Indeed, we could have lived in a world where Franklin and Church's frameworks dominated instead of Watson-Crick and Turing, yet science would still have advanced. However, while breakthroughs may be inevitable, funding, infrastructure, and collaboration within the scientific ecosystem remain essential—not to ensure breakthrough happen, but to shape how and when they do.

**Data availability**

The datasets for reproducing the main findings of the D-index are available on Harvard Dataverse: https://doi.org/10.7910/DVN/VE3AFX.

# Supplementary Materials for
"Is Science Inevitable? Large-Scale Data Reveals That Breakthroughs Are Multiple, Not Singular"

Linzhuo Li, Yiling Lin and Lingfei Wu

Corresponding author: linzhuoli@zju.edu.cn (L.L) and liw105@pitt.edu (L.W)

**The PDF file includes:**

Materials and Methods
Figures S1 to S7
Table S1

Materials and Methods

These supplementary materials provide:
1. A description of the three major datasets.
2. An introduction to the D-index.
3. A description of the survey on breakthrough papers.
4. Use of Large Language Models to identify theoretical and methodological breakthroughs.

S1. Description of three major datasets

We created and analyzed four datasets:

1. **Breakthrough Papers Dataset**: This dataset includes 105 breakthrough papers nominated by a panel of twenty scientists from leading research institutions across five countries and nine disciplines. We linked these papers to our other datasets and retrieved their research metrics (see Table S1 for the ten most recommended papers).

2. **Microsoft Academic Graph (MAG) Dataset**: This dataset includes the publicly available version of the Microsoft Academic Graph (MAG), which can be downloaded at https://zenodo.org/records/2628216. We focused on papers published between 1965 and 2020. Our MAG dataset includes 40,935,251 journal articles (excluding books and conference papers, as citations to them largely follow different scientific practices).

3. **Multiple-version Papers Dataset**: This subset of MAG contains papers with multiple published versions, typically including initial preprints in digital archives and later versions published in peer-reviewed journals. This dataset provides an unprecedented opportunity for quasi-experimental analysis, allowing us to examine the marginal effect of changes in paper content on citation impacts for the same paper. It includes 3,382 papers with two versions, with an average temporal separation of 1.84 years between the first and the final versions.

4. **Funded Papers Dataset:** This dataset is a subset of "SciSciNet," an open dataset (https://www.nature.com/articles/s41597-023-02198-9), which extends from MAG and incorporates various other sources, including published papers acknowledging support from the National Science Foundation (NSF) and the National Institutes of Health (NIH). We linked 2,357,593 NIH-funded papers and 652,652 NSF-funded papers (1980-2020) with our MAG dataset and examined the innovation metrics of the papers supported by these federal agencies.

S2. Explanation of two research metrics: D-index and A-index

S2.1 Definition of *D*-index

Calculating the D-index involves dividing subsequent papers citing a focal paper $p$ into three types:
1. **Type $i$:** Papers that exclusively cite the focal paper but disregard its references.
2. **Type $j$:** Papers that cite both the focal paper and its references.
3. **Type $k$:** Papers that solely cite the references of the focal paper but not the focal paper itself.

The D-index is defined as the disparity between type $i$ and type $j$ papers, normalized by the sum of these three types of papers. Figure S1 provides a simplified illustration.

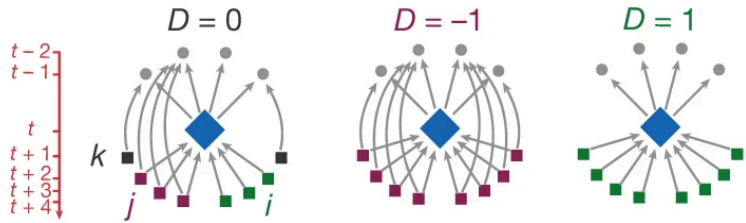

**Figure S1. Simplified illustration of *D*-index.** The figure is reproduced from our earlier research [8].

The formula is as follows, with $N$ denoting the number of each type of paper:

$$D_f = \frac{N_i - N_j}{N_i + N_j + N_k} \qquad \text{Eq. (1)}$$

S2.2 What does the *D*-index measure?

The Disruption (*D* or *CD*) index, which technically measures how subsequent papers cite a focal paper while disregarding its references, has been used to identify scientific breakthroughs [4,8]. Despite its great potential, previous studies have struggled to understand its theoretical implications [9,15]. Through years of investigation, we find that the D-index reflects how a new idea displaces old ones while addressing similar questions. This is based on two insights:

(a) Displacing the most cited reference

First, $D$ reflects the substitutive relationship between the focal paper and its most cited reference, rather than the absolute innovation level of the focal paper as commonly believed [4,8]. This is because to achieve a high D-index, a paper must compete against its references for future citations, primarily against the most cited reference due to the long-tail distribution of citation impacts among these references. We can see this insight by rearranging Eq. 1 [16]:

$$D_f = \frac{(N_i - N_j)/(N_i + N_j)}{1 + N_k/(N_i + N_j)} = \frac{1}{1+R_k} d_f \qquad \text{Eq. (2)}$$

Here we decompose the D-index into two terms, $d_f$ and $R_k$. $d_f = (N_i - N_j)/(N_i + N_j)$ is a "local" measure reflecting the focal paper $f$'s intrinsic innovative level based on the two types of citing papers of it. The other term, $R_k = N_k/(N_i + N_j)$ can be approximated as $C_{total}/(N_i + N_j)$, where $N_k$ (the exclusive citations to the references) serves as a proxy for $C_{total}$, the total citations to the references. This approximation holds because shared citations between the focal paper and its references are generally smaller than total reference citations. As a result, $R_k$ quantifies citation competition between the focal paper and its references, decreasing when the focal paper gains more citations than its references.

Notably, this competition primarily occurs between the focal paper and its most cited reference, as reference citations follow Zipf's law:

$$C_r \propto c \frac{1}{(b+r)^a} \qquad \text{Eq. (3)}$$

where $c$ is a constant, rank $r$ represents the rank of the reference in descending citation order, and $C_r$ denotes the citation impact of the reference ranked $r$. Parameter $a$ indicates how unequal the citation distribution is and parameter $b$ as a fitted constant. Figure S2a illustrates the relationship between reference ranks and their corresponding citation counts, along with the fitted Zipf curve.

To quantify the dominance of the most cited reference ($r=1$) over the total citations, we calculate the ratio of the citations of the most cited reference to the total citations:

$$C_{max}/C_{total} = c\frac{1}{(b+1)^a} \Big/ \sum_{r=1}^{N} c\frac{1}{(b+r)^a} \qquad \text{Eq. (4)}$$

where $C_{max}$ represents the citation count of the most cited reference, and $C_{total}$ is the total number of citations across all references. From Eq. 4, for large $N$ and small $b$, the total number of citations can be expressed as:

$$\begin{aligned}
C_{total} &= \sum_{r=1}^{N} c\frac{1}{(b+r)^a} \\
&\approx c \int_{1}^{N} \frac{dx}{(b+x)^a} \\
&\approx c\left(\frac{(b+N)^{1-a}}{1-a} - \frac{(b+1)^{1-a}}{1-a}\right)
\end{aligned} \qquad \text{Eq. (5)}$$

Thus, the ratio $C_{max}/C_{total}$ simplifies to:

$$C_{max}/C_{total} \approx \frac{1}{(b+1)^a} \Big/ \left(\frac{(b+N)^{1-a}}{1-a} - \frac{(b+1)^{1-a}}{1-a}\right) \qquad \text{Eq. (6)}$$

Although the reference length $N$ varies across papers (Figure S2b), the term $b+N$ is one order of magnitude larger than $b+1$ ($b+1$ is on the order of $10^0$, while $b+N$ is on the order of $10^1$). Additionally, the highly skewed citation distribution ($a>1$, Figure S2c) ensures that $(b+N)^{1-a}$ is negligible compared to $(b+1)^{1-a}$. Therefore, $N$ has minimal impact on the ratio $C_{max}/C_{total}$:

$$C_{max}/C_{total} \approx \frac{1}{(b+1)^a} \Big/ \frac{(b+1)^{1-a}}{a-1} \approx \frac{a-1}{1+b} \qquad \text{Eq. (7)}$$

Using the average parameter values $a=2.0$ (Figure S2c) and $b=1.4$ (Figure S2d), we estimate:

$$C_{max}/C_{total} \approx \frac{2-1}{1+1.4} = \frac{1}{2.4} = 0.42 \qquad \text{Eq. (8)}$$

As a result, the theoretical value of $C_{max}/C_{total}$, computed from Eq. 8, is 0.42, which closely matches the empirical value of 0.40 (Figure S2e).

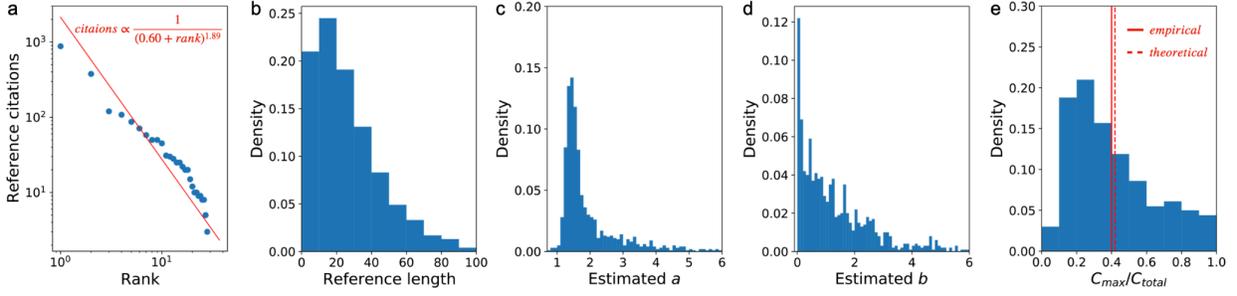

**Figure S2. Zipf's law of paper reference citations and the characteristics value of $C_{max}/C_{total}$.** We randomly selected 1,000 journal papers from 1900 to 2020 with three or more references. This reference length threshold ensures accurate parameter estimation. Panel (a) illustrates the citations of references cited by a paper against their rank in descending order (blue dots) with the Zipf's law fit (red line). The estimated parameters are $a = 1.89$ and $b = 0.60$. Panel (b) displays the distribution of the reference lengths, with an average of 29. Panel (c) displays the distribution of the estimated $a$ values for the 1,000 papers; 99.3% of them are all greater than 1, with an average of 2. The ratio $C_{max}/C_{total}$ quickly converges to a constant $(b+1)/(a-1)$ when the reference length $N$ increases and becomes independent of the reference length $N$ when $a>1$. Panel (d) shows the distribution of the estimated values of b, with an average of 1.4. Panel (e) displays the distribution of $C_{max}/C_{total}$, with an average of 0.40, as shown in the red solid line. The theoretical value, as shown in the red dashed line, is calculated using Eq. 8, is 0.42. The empirical value of $C_{max}/C_{total}$ closely matches the theoretical value.

The empirical results suggest that $C_{total}$ can be approximated as $2.5C_{max}$. Thus, If we define $C_f = N_i + N_j$ as the total citations to the focal paper, we can rewrite Eq. 2 as:

$$D_f = \frac{d_f}{1+R_k} \approx \frac{d_f}{1+C_{total}/C_f} \approx \frac{d_f}{1+2.5C_{max}/C_f} \qquad \text{Eq. (9)}$$

or:

$$D_f \approx \frac{1}{1+b_f}d_f \qquad \text{Eq. (10)}$$

where $b_f \sim C_{max}/C_f$. Eq. 10 reveals that the D-index is primarily determined by two variables: the local displacement factor ($d_f$) and the "burden" factor ($b_f$).

**(1) The local displacement factor, $d_f = N_i/C_f - N_j/C_f = p_i - p_j$**, measures the disparity between the probability of observing two types of subsequent papers citing a focal paper. It reflects the relationship between the focal paper and its top reference, as determined by the offspring papers that cite the focal paper. If $d_f > 0$, the focal paper undermines the influence of its top reference; if $d_f < 0$, it consolidates and enhances the reference's influence; if $d_f = 0$, it is neutral. Most papers in our dataset have $d_f < 0$ (Figure S3a).

**(2) The knowledge burden factor, $b_f = C_{max}/C_f$**, represents the ratio of the top reference's citation impact to that of the focal paper, based on the "burden of knowledge" theory [10]. It characterizes the global impact of the focal paper's role in displacing or consolidating the top reference. $D_f$ is always smaller than $d_f$ because $D_f$ equals $d_f$ multiplied by a factor smaller than one: $1/(1+b_f)$. When the burden factor is large ($b_f > 1$), $D_f$ approximates zero, indicating that the focal paper has no impact on its top reference. Conversely, when the burden factor is small (0 <

$b_f < 1$), $D_f$ approximates $d_f$, meaning the focal paper's potential to displace or consolidate the top reference is fully realized. Most papers in our dataset have $b_f > 1$ (Figure S3b).

Combining these two factors explains why the majority of papers have a D-index close to zero (Figure S3c). This is not because they lack the potential to displace or consolidate their top reference, but because this potential is recognized locally, not globally. While great papers often "stand on the shoulders of giants," as Newton said, not all papers standing on the shoulders of giants will become giants. Only those who shed the burden of knowledge from previous giants will become new giants and successfully displace their predecessors.

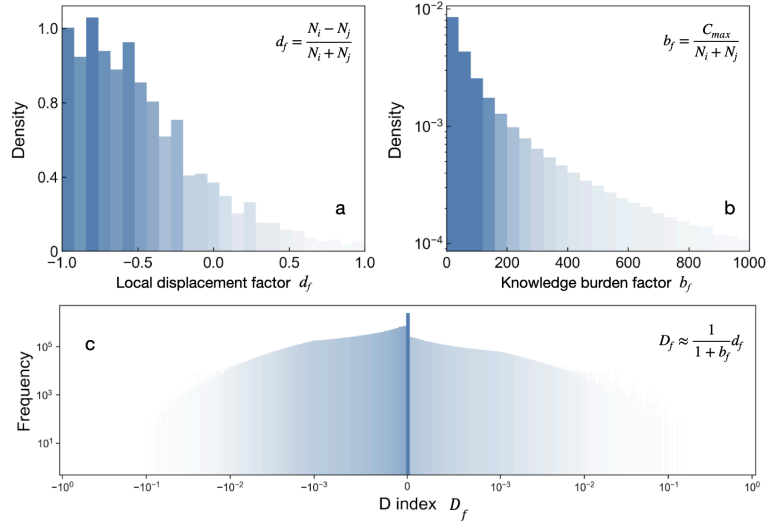

**Figure S3. The distribution of $d_f$ and $b_f$.** Panels (a) and (b) show the distribution of the local displacement factor ($d_f$) and the knowledge burden factor ($b_f$) for 3,164,274 journal papers (1965-2020) with ten or more citations. This citation impact threshold allows us to focus on the continuous change in density. If citations are too few, these two variables, being ratios of natural numbers, will only take a few possible values, making it hard to observe a continuous change. Among these papers with ten or more citations, 85% have negative $d_f$, 13% have positive $d_f$, and 2% have a $d_f = 0$. For $b_f$ (Panel b), 99.7% of papers have $b_f > 1$, 0.2% have $b_f < 1$, and 0.01% have $b_f = 1$. The dominance of negative $d_f$ (74.5%) and $b_f$ greater than one (99.8%) remains highly consistent when we include all papers with one or more citations. Panel (c) shows the distribution of the D-index ($D_f$) across 40,935,251 journal articles (1965-2020) with one or more citations and references. 30% of these papers have a positive $D_f$, while 70% have zero or negative $D_f$.

(b) Displacing within the same research topic

Second, high-D papers and their top references tend to overlap in topics, with a probability much higher than predicted by random models based on recombinant growth theory [11]. In addition to the examples in the main text, Figure S4 below presents three examples from our Breakthrough Papers Dataset, highlighting the significant topic overlap between the focal papers and their top references. The complete list of all ten cases is shown in Table S1.

| Topics | Breakthrough Papers (Upper) and Their Most Impactful References (Lower) | D |
|---|---|---|
| Biology, DNA structure 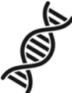 | Watson, J. D., & Crick, F. H. (1953). Molecular structure of nucleic acids. Nature, 171(4356), 737-738. | 0.88 |
| | Pauling, L., & Corey, R. B. (1953). A proposed structure for the nucleic acids. Proceedings of the National Academy of Sciences of the United States of America, 39(2), 84. | |
| Physics, Fluid dynamics 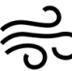 | Lorenz,E.N.(1963). Deterministic nonperiodic flow. Journal of the atmospheric sciences, 20(2), 130-141. | 0.86 |
| | Rayleigh, L. (1916). LIX. On convection currents in a horizontal layer of fluid, when the higher temperature is on the under side. The London, Edinburgh, and Dublin Philosophical Magazine and Journal of Science, 32(192), 529-546. | |
| Mathematics, Fractal geometry 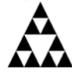 | Mandelbrot, B. (1967). How long is the coast of Britain? Statistical self-similarity and fractional dimension. science, 156(3775), 636-638. | 0.83 |
| | Steinhaus, H. (1954). Length, shape and area. In Colloq. Math. (Vol. 3, pp. 1-13). | |

**Figure S4. Examples of three displacing papers.** The full list of ten papers is displayed in Table S1.

We quantified how often displacing papers ($D > 0$) and their top references share overlapping topics. Leveraging the scientific taxonomy MAG used to label papers [17], we assigned each paper in our MAG dataset to 292 fields, such as Discrete Mathematics, Molecular Biology, and Organic Chemistry. We then built a random model on field selection to compare the empirical against the theoretical probability of overlap between a displacing paper and its most impactful reference. On average, a paper has two unique field labels. If a paper randomly selects two different field labels, the probability of two random papers having overlapping labels is:

$$p = 1 - \frac{\binom{290}{2}}{\binom{292}{2}} \approx 0.015 \qquad \text{Eq. (11)}$$

The empirical probability we observed was $p = 0.62$, which is 40 times higher than this theoretical prediction. The high topic alignment between displacing papers and their top references suggests that breakthroughs often occur as purposeful innovations, rather than merely from the random recombination of prior knowledge.

In conclusion, we suggest that the *D*-index measures how a new idea displaces old ones while addressing similar questions or phenomena [18]. It highlights the relationship between two papers rather than just assessing the inventive level of a single paper, thus adding focus to the historical trajectories of science rather than interrupting or breaking them. Therefore, while Disruption remains an appealing name, we might also consider interpreting the D-index as the Displacement Index. We hope this clarification will better communicate the nature of this innovation measure and promote its more accurate use in research evaluation [9].

S2.3 Technical issues in calculating the *D*-index

There is an ongoing debate about the D-index's technical complexity [19–22], raising valid concerns about its patterns and interpretations [4]. Guided by our new understanding of what the D-index actually measures—the displacing relationship between a focal paper and its top reference—we will discuss recent concerns aimed at making optimal decisions on these technical issues. We note that these discussions do not represent final decisions and call for further studies to enhance the accuracy and usefulness of the D-index as an innovation measure.

(a) D-index and reference length

As science advances, more papers are published every year, and each paper cites more prior papers. This phenomenon is called "citation inflation" due to its similarity with monetary inflation caused by an increase in the money supply in economic systems [23]. A recent study raised valid concerns about whether this could confound the observed decline in the average value of the D-index of all papers [4], and make the temporal analysis of the D-index challenging in general [24]. The rationale is that the more references a focal paper has, the less likely it will have a high D-index due to the increased denominator in Eq. 1. In theory, the D-index may converge to zero as citations continue to inflate.

While these studies reveal the increased knowledge burden to all papers imposed by lengthening references and its impact on the D-index, we have shown that the true burden is less about the number of references and more about the largest value of their citation impacts, following Zipf's law. The citation impact of the top reference that the focal paper displaces (or consolidates) determines the task's difficulty, and the consequential "burden factor", the ratio of citation impacts between the top reference and the focal paper, measures the extent to which the focal paper successfully affect the role of its top reference in the scientific field. In other words, the burden comes from which "giant" in the canonical literature the focal paper chooses to challenge, not the time period the focal paper is born into.

To illustrate this point, we show that the D-index is independent of reference length after accounting for the local displacement factor ($d_f$) and the burden factor ($b_f$). See Figure S5.

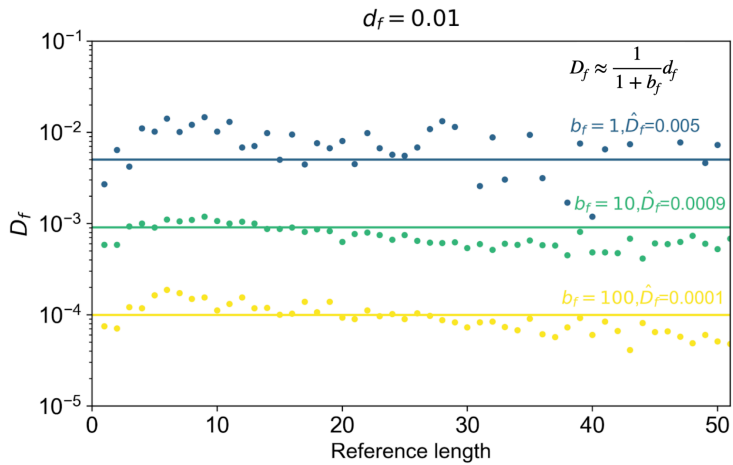

**Figure S5. D-index is independent of reference length after accounting for $d_f$ and $b_f$.** We select 606,081 papers with an average local displacing index ($d_f$) of 0.01, with values ranging from 0 to 0.112. We then further divided them into subgroups based on the burden factor, including $b_f=1$ (664 papers), $b_f=10$ (22,829 papers), and $b_f=100$ (52,857 papers). The empirical values of the D-index for these papers align with their theoretical predictions given by Eq. 10, as shown in the figure.

We acknowledge that the analysis here is based on approximations. In practice, to fully exclude the direct impact of reference length, one may focus on the sign of $D_f$ rather than its average value for quality judgment and temporal analysis [16]. In the main text, we follow this rationale and calculate the fraction of papers with $D_f>0$ to investigate the temporal change of displacing papers.

(b) The discriminative power of the D-index and the development of alternatives

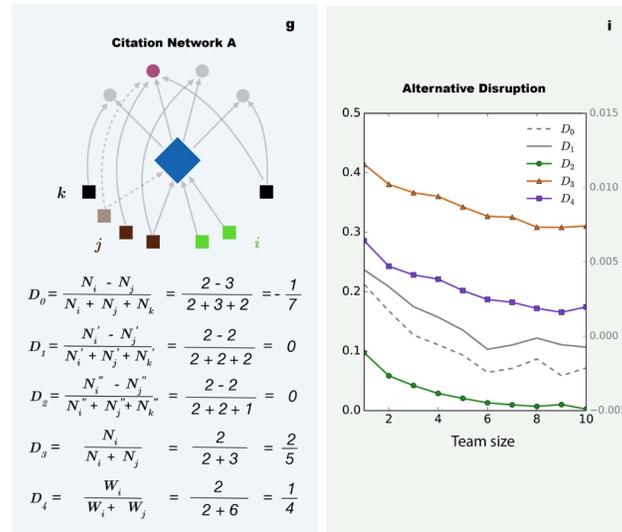

**Figure S6. Alternative versions of *D*-index.** The figure is reproduced from our earlier research [8]. Panel (g) shows a simplified citation network comprising focal papers (diamonds), references (circles), and subsequent work (rectangles). Subsequent work may cite (1) only the focal work (*i*, green), (2) only its references (*k*, black), or (3) both focal work and references (*j*, brown). A reference identified as popular is colored in red, and self-citations are shown by dashed lines (with corresponding subsequent work colored in light brown). Five definitions of the D-index are provided for comparison. $D_0$ is the definition used in the main text. $D_1$ is defined the same way as $D_0$, but with self-citations excluded. $D_2$ is defined the same way as $D_0$ but only considers popular references. In the empirical analysis, we identified references as popular that received citations within the top quartile of the total citation distribution (≥24 citations). $D_3$ simplifies $D_0$ by only measuring the fraction of papers that cite the focal paper and not its references, among all papers citing the focal paper. $D_4$ is similar to $D_3$ but weighted by the number of citations. For example, if a single referenced paper is cited five times, then it receives a count of five rather than one. Panel (i) shows that all alternative measures to the D-index decrease consistently with team size. $D_0$ and $D_1$ are indexed by the right y-axis and other disruption measures are indexed by the left y-axis. One hundred thousand randomly selected Web of Science papers (97,188 papers remained after excluding missing data) are used to calculate these values.

Recent research has raised concerns about the D-index's discriminative power, particularly because its numerator is bounded while its denominator is unbounded (Eq. 1). This formulation tends to produce values close to zero, potentially undermining its discriminative power [16]. To address this issue, some studies have explored alternative versions of the D-index [25]. We would like to highlight the value of the current D-index for two reasons.

$$D_f \approx \frac{1}{1+b_f} d_f \qquad \text{Eq. (10)}$$

First, the D-index ($D_f$) is actually highly effective in identifying revolutionary work. Our decomposition (Eq. 10) of the D-index into a local displacement factor ($d_f$) and a knowledge burden factor ($b_f$) explains why the D-index tends to be zero: most papers cite canonical literature with much larger citation impacts. While some scholars see this as a limitation, it actually highlights a small set of papers whose role in displacing or consolidating their top reference is significant, distinguishing them from the majority of papers with negligible impact compared to their top reference.

For example, given a local displacement factor $d_f = 0.5$, maintaining this displacing effect globally is increasingly difficult. If the focal paper has the same citation impact as its top reference ($b_f=1$), which occurs in only 0.2% of cases based on Figure S3, $D_f$ is reduced to 0.5/2= 0.25. If the focal paper has twice the citation impact of its top reference ($b_f=0.5$), $D_f$ reduces to 0.5/1.5=0.33. If the focal paper has ten times the citation impact of its top reference ($b_f=0.1$), $D_f$ reduces to 0.5/1.1=0.45. Such occurrences are extremely rare and may only happen a couple of times in a scientific field within a decade, especially if the top reference is canonical literature. In other words, when a highly positive D-index is observed, it means the idea presented by the paper not only substitutes its top reference but is also well-recognized by the field

Therefore, if the goal is to identify revolutionary work in the history of science, the global D-index ($D_f$) has a higher discriminative power. If the goal is to distinguish between normal papers, the local displacement factor ($d_f$) is more effective.

Second, alternative versions of the D-index seem to exhibit similar behaviors. For example, a recent study stated that "the results of a factor analysis show that the different variants measure similar dimensions" [25]. Our previous research [8] has also considered five versions of the D-index, all of which demonstrated consistent correlation with another variable, team size (Figure S6). This includes a version ($D_3$) that has excluded self-citations, which may, therefore, ease the concern raised in [16].

Additionally, recent research proposed that the original D-index may not be accurate and should be weighted by the number of citations to account for both the magnitude and reach of high D-index papers [21]. This is similar to the weighted version ($D_4$) presented in Figure S6. To fully address this concern, we explore a weighted version following the formulation of [21] and observed the decline of displacing papers from 60% in 1965 to 20% in 2020, exhibiting an even stronger trend than shown in the main text (from 50% to 20%).

(c) Data requirements in calculating the *D*-index

A recent study raised concerns about missing data affecting the calculation of the D-index [19]. Accurate D-index calculation relies on the complete retrieval of paper references. Missing values in paper references can distort the D-index. For example, a paper with references but no citations results in D=0. This can be misleading as papers with many citations but evenly split between subsequent papers that also cite its reference or disregarding them also result in D = 0. Similarly, papers with citations but no references may misleadingly result in D=1. This can distort the explanation of results, as our previous analysis shows that achieving a high positive D-index is very difficult and a rare event in the real world.

These issues can affect the D-index itself and skew downstream analyses correlating D with other variables, especially if missing data is unevenly distributed. To ensure the D-index is as reliable as possible, we include only papers with one or more references and citations in our analysis in this manuscript and previous studies [8,15,26].

After all, the D-index measures intellectual contribution based on citation practices, reflecting how a focal work relates to preceding major ideas as determined by the subsequent papers. Without reference or citation data, this analysis is not possible.

We also recommend focusing on one type of publication venue at a time, rather than combining journals, conferences, theses, books, or essays, which have different citation practices and could affect D-index interpretation. In our previous studies, we typically focus on journal articles to minimize issues from varying citation practices. Citation practices are more established for academic journals, and the peer-review mechanism further serves as quality control for these norms.

(d) D-index and betweenness centrality

Recent literature suggests that the *D*-index is a specific form of node centrality in citation networks: betweenness [27]. Betweenness centrality measures how often a node appears on the shortest paths between other nodes, indicating its role as a bridge within the network [28]. While we agree with this topological interpretation, we would like to emphasize that it should not confuse the originality and meaning of the *D*-index.

    First, using node centrality to measure paper importance has a long history [29], inspiring the PageRank algorithm in information retrieval [30]. However, to our knowledge, these network measures usually focus on the topology of the citation network and rarely leverage the hidden time dimension as the D-index does. The D-index is unique in considering this temporal aspect, highlighting papers with high D-index as "gatekeepers in time."

    Second, Interpreting the *D*-index as merely betweenness centrality in networks risks focusing on the strategic advantage of high *D*-index papers as "knowledge brokers" and ignoring their inherent intellectual contributions. While being a knowledge broker in social networks often reflects social capital advantages [31,32], achieving this status in citation networks is hard-earned. For example, in our Breakthrough Papers Dataset, the 1998 small-world paper by Duncan Watts and Steve Strogatz [33] displaced Stanley Milgram's 1967 paper [34] (see Table S1). It is an oversimplification to assume that subsequent citations of Watts and Strogatz were simply due to ignorance of Milgram's work, as scientists are trained to discover and cite the original literature. Based on our interview, the high *D*-index of Watts and Strogatz correctly reflects its radical advancement from Milgram's work, by providing a novel mathematical framework to quantify the small-world phenomenon beyond social networks.

(e) The negative values of the D-index

Regarding the interpretation of the D-index, it is important to note that papers with a negative D can also significantly contribute to science, as described by the term "consolidation," hence the name "CD-index" [7]. For example, Wolfgang Ketterle et al.'s paper on Bose-Einstein Condensation [35], which validated the theory proposed by Albert Einstein and Satyendra Nath Bose through lab experiments has a D=-0.017 (bottom 6% among all papers) [8]. Ketterle won the Nobel Prize in Physics in 2001 for this work. Another example is the 2001 human genome paper [36], an important milestone in genomic research resulting from massive international collaboration. Both of these works consolidate revolutionary scientific ideas—the Einstein-Bose condensation theory and the DNA structure—rather than displacing them.

    The asymmetric distribution of the D-index (70% D ≤ 0 based on Figure S3c) suggests that consolidating innovation is the norm in science. This pattern differs from that in technology, where more patents have a positive D-index (62%) than a negative one (38%), based on

open-source data we published [26]. It would be interesting to explore whether this reflects fundamental differences in the level of path dependency between science and technology.

S3. Description of the Survey on Breakthrough Papers

In 2019, we conducted an open-ended survey on identifying breakthrough research in science, performed in person, over the telephone, or using Skype, approved by the University of Chicago Institutional Review Board (IRB18-1248). The survey asked scholars across various fields to propose papers that either disrupt or develop science in their fields, using the following definitions: (a) Developing papers: Extensions or improvements of previous theory, methods, or findings; (b) Disrupting papers: Punctuated advances beyond previous theory, methods, or findings. Note that the D-index was referred to as "Disruption" at the time, so we use this term hereafter to remain consistent with the survey.

We provided respondents with examples like the BTW model [37] and Bose-Einstein condensation [35] papers to illustrate disruptive and developmental papers. Respondents then proposed three to ten disrupting and developing papers. Our panel included scientists from ten prominent research-intensive institutions across the United States, China, Japan, France, and Germany, with training in mathematics, physics, chemistry, biology, medicine, engineering, computer science, psychology, and economics.

Among the 20 scholars from whom we received 190 responses, 100% of their proposals for the most disruptive paper agreed with our measure, and all but six proposals for the most developing paper agreed with our measure. The average disruption score of papers nominated as disruptive is $D=0.2147$, placing them in the top 2% of most disruptive papers. The average disruption score of papers nominated as developing is $D=-0.011$, placing them in the bottom 13%. This analysis resulted in an overall prediction area under the curve of 0.83, suggesting a predictive accuracy of 83% and a strong sensitivity to extremes.

For the current research, we re-analyzed the data to select the top ten nominated papers from the survey and identified their most cited references, as presented in Table S1.

Besides the survey, we analyzed nine breakthrough papers selected by Nature editors to celebrate the journal's 150th anniversary. Notably, Watson and Crick's 1953 paper on the structure of DNA appears in both tables, reflecting a broad consensus on its breakthrough significance. The selected papers have an average disruption score of $D = 0.48$, placing them in the top 1% of the most disruptive papers. We also identified their most cited references, presented in Table S2.

**Table S1. Top Ten Nominated Breakthrough Papers.** These breakthroughs were selected as the most recommended papers from an open-ended survey conducted with twenty scientists from prominent research institutions spanning five countries and nine disciplines. For each breakthrough paper, its most cited references are manually retrieved by human coders using Google Scholar.

|   | Topics | Breakthrough Papers and Their Most Cited References | *D*-index |
|---|---|---|---|
| 1 | Biology, DNA structure | Watson, J. D., & Crick, F. H. (1953). Molecular structure of nucleic acids. Nature, 171(4356), 737-738. | 0.88 |
|   |   | Pauling, L., & Corey, R. B. (1953). A proposed structure for the nucleic acids. Proceedings of the National Academy of Sciences of the United States of America, 39(2), 84. |   |

| 2 | Physics, Fluid dynamics | Lorenz,E.N.(1963). Deterministic nonperiodic flow. Journal of the atmospheric sciences, 20(2), 130-141. | 0.86 |
|---|---|---|---|
| | | Rayleigh, L. (1916). LIX. On convection currents in a horizontal layer of fluid, when the higher temperature is on the under side. The London, Edinburgh, and Dublin Philosophical Magazine and Journal of Science, 32(192), 529-546. | |
| 3 | Mathematics, Fractal geometry | Mandelbrot, B. (1967). How long is the coast of Britain? Statistical self-similarity and fractional dimension. science, 156(3775), 636-638. | 0.83 |
| | | Steinhaus, H. (1954). Length, shape and area. In Colloq. Math. (Vol. 3, pp. 1-13). | |
| 4 | Sociology, Small-world dynamics | Watts, D. J., & Strogatz, S. H. (1998). Collective dynamics of 'small-world' networks. nature, 393(6684), 440. | 0.48 |
| | | Milgram, S. (1967). The small world problem. Psychology today, 2(1), 60-67. | |
| 5 | Computer science, Text classification method | Blei, D. M., Ng, A. Y., & Jordan, M. I. (2003). Latent dirichlet allocation. Journal of machine Learning research, 3(Jan), 993-1022. | 0.43 |
| | | Nigam, K., McCallum, A. K., Thrun, S., & Mitchell, T. (2000). Text classification from labeled and unlabeled documents using EM. Machine learning, 39(2), 103-134. | |
| 6 | Psychology, Number sense theory | Wynn, K. (1992). Addition and subtraction by human infants. Nature, 358(6389), 749. | 0.36 |
| | | Starkey, P., & Cooper Jr, R. G. (1980). Perception of numbers by human infants. Science, 210(4473), 1033-1035. | |
| 7 | Economics, Game theory | Nash,J. (1951). Non-cooperative games. Annals of mathematics, 286-295. | 0.28 |
| | | Von Neumann, J, & Morgenstern. O. (1944) Theory of games and economic behavior. Princeton university press | |
| 8 | Computer science, Computability theory | Turing, A. M. (1937). On computable numbers, with an application to the Entscheidungs problem. Proceedings of the London Mathematical Society, 2(1), 230-265. | 0.22 |

| | | Gödel, K. (1931). On formally undecidable theorems of Principia Mathematica and related systems I. Monthly journal for mathematics and physics , 38 , 173-198. | |
|---|---|---|---|
| 9 | Engineering, Optimization algorithms | Kirkpatrick, S., Gelatt, C. D., & Vecchi, M. P. (1983). Optimization by simulated annealing. science, 220(4598), 671-680. | 0.20 |
| | | Metropolis, N., Rosenbluth, A. W., Rosenbluth, M. N., Teller, A. H., & Teller, E. (1953). Equation of state calculations by fast computing machines. The journal of chemical physics, 21(6), 1087-1092. | |
| 10 | Computer Science, Neural attention mechanisms | Vaswani, A., Shazeer, N., Parmar, N., Uszkoreit, J., Jones, L., Gomez, A. N., ... & Polosukhin, I. (2017). Attention is all you need. Advances in neural information processing systems, 30. | 0.03 |
| | | Hochreiter, S., & Schmidhuber, J. (1997). Long short-term memory. Neural computation, 9(8), 1735-1780. | |

**Table S2. Breakthrough Papers from Nature's 150th Anniversary Collection.** These papers were selected by Nature for its 150th anniversary. For each breakthrough paper, its most cited references were manually retrieved by human coders using Google Scholar.

| | Topics | Breakthrough Papers and Their Most Cited References | *D*-index |
|---|---|---|---|
| 1 | Biology, DNA structure | Watson, J. D., & Crick, F. H. (1953). Molecular structure of nucleic acids. Nature, 171(4356), 737-738. | 0.88 |
| | | Pauling, L., & Corey, R. B. (1953). A proposed structure for the nucleic acids. Proceedings of the National Academy of Sciences of the United States of America, 39(2), 84. | |
| 2 | Chemistry, Carbon cluster structure | Kroto, H. W., Heath, J. R., O'Brien, S. C., Curl, R. F., & Smalley, R. E. (1985). C60: Buckminsterfullerene. Nature, 318(6042), 162-163. | 0.80 |
| | | Rohlfing, E. A., Cox, D. M., & Kaldor, A. (1984). Production and characterization of supersonic carbon cluster beams. The Journal of chemical physics, 81(7), 3322-3330. | |
| 3 | Environmental Science, Stratospheric processes | Farman, J. C., Gardiner, B. G., & Shanklin, J. D. (1985). Large losses of total ozone in Antarctica reveal seasonal ClO x/NO x interaction. Nature, 315(6016), 207-210. | 0.76 |
| | | Dunkerton, T. (1978). On the mean meridional mass motions of the stratosphere and mesosphere. Journal of the Atmospheric Sciences, 35(12), 2325-2333. | |

| 4 | Biology, Antibody production | Köhler, G., & Milstein, C. (1975). Continuous cultures of fused cells secreting antibody of predefined specificity. Nature, 256(5517), 495-497. | 0.69 |
| --- | --- | --- | --- |
| | | Jerne, N. K., & Nordin, A. A. (1963). Plaque formation in agar by single antibody-producing cells. Science, 140(3565), 405-405. | |
| 5 | Neuroscience, Ion channel currents | Neher, E., & Sakmann, B. (1976). Single-channel currents recorded from membrane of denervated frog muscle fibres. Nature, 260(5554), 799-802. | 0.37 |
| | | Anderson, C. R., & Stevens, C. F. (1973). Voltage clamp analysis of acetylcholine produced end‑plate current fluctuations at frog neuromuscular junction. The Journal of physiology, 235(3), 655-691. | |
| 6 | Physics, Particle detection | ROCHESTERDr, G. D., & BUTLERDr, C. C. (1947). Evidence for the existence of new unstable elementary particles. Nature, 160(4077), 855-857. | 0.36 |
| | | Lattes, C. M. G., Occhialini, G. P., & Powell, C. F. (1947). Observations on the tracks of slow mesons in photographic emulsions. Nature, 160(4067), 486-492. | |
| 7 | Materials Science, Porous material synthesis | Kresge, A. C., Leonowicz, M. E., Roth, W. J., Vartuli, J. C., & Beck, J. S. (1992). Ordered mesoporous molecular sieves synthesized by a liquid-crystal template mechanism. Nature, 359(6397), 710-712. | 0.21 |
| | | Gregg, S. J., Sing, K. S. W., & Salzberg, H. W. (1967). Adsorption surface area and porosity. Journal of The electrochemical society, 114(11), 279Ca. | |
| 8 | Astronomy, Stellar companions | Mayor, M., & Queloz, D. (1995). A Jupiter-mass companion to a solar-type star. Nature, 378(6555), 355-359. | 0.21 |
| | | Duquennoy, A., & Mayor, M. (1991). Multiplicity among solar-type stars in the solar neighbourhood. II-Distribution of the orbital elements in an unbiased sample. Astronomy and Astrophysics (ISSN 0004-6361), vol. 248, no. 2, Aug. 1991, p. 485-524. Research supported by SNSF., 248, 485-524. | |
| 9 | Biology, Nuclear reprogramming | Gurdon, J. B., Elsdale, T. R., & Fischberg, M. (1958). Sexually mature individuals of Xenopus laevis from the transplantation of single somatic nuclei. Nature, 182(4627), 64-65. | 0.08 |
| | | King, T. J., & Briggs, R. (1956, January). Serial transplantation of embryonic nuclei. In Cold Spring Harbor symposia on quantitative biology (Vol. 21, pp. 271-290). Cold Spring Harbor Laboratory Press. | |

S4. Using Large Language Models to distinguish theoretical and methodological breakthroughs

For all 50,147 high impact, disruptive(D>0) papers in our dataset, we employ Llama-3.1-70B-Instruct, the state-of-art open-sourced Large Language Model published by Meta, to classify the type of breakthroughs each paper has relative to its top-cited reference. Specifically, we utilized Llama-3.1-70B-Instruct to analyze pairs of academic papers, consisting of a primary paper and its most cited reference. The model was presented with the titles and abstracts of both papers and asked to classify the primary paper's innovative contribution using the following zero-shot prompt:

*"Given two papers - Paper A {title of paper} with abstract {paper abstract}, and its primary reference Paper B {title of reference} with abstract {reference abstract} - Paper A is better considered as an innovation in 'theory' (such as a different conceptual difference from its reference paper above), (2): an innovation in 'method' (such as an improvement on or formalization of the latter in mathematical frameworks)? Only give the option number."*

As probabilistic models, Large Language Models (LLMs) generate predictions through multinomial distributions over possible next tokens. When predicting an answer (e.g., "1"), the model assigns probabilities to all potential choices ("1" and "2" in this case) before sampling from this distribution. This probabilistic framework enables us to directly analyze the model's confidence in its classification without requiring multiple sampling iterations. Instead, we can examine the token probability distribution at the prediction point to determine the relative likelihood of each classification option. By extracting these probabilities, we calculate $p$, representing the model's confidence in classifying a paper as a theoretical breakthrough.

We use the ten breakthrough examples from our survey as a quick validation set of LLM codings (Figure S7). As can be seen, the Llama model was able to make distinguishing choices for theory and method innovations. For example, in analyzing Turing and Church's papers, the model assigned $p = 0.86$ to option (1). While for the Watts and Strogatz paper, $p = 0.63$. Another interesting example is Nash's paper of game theory ($p=0.9$), which offers a distinct conceptual approach of non-cooperative game theory from the cooperative game theory in Von Neumann and Morgenstern's book.

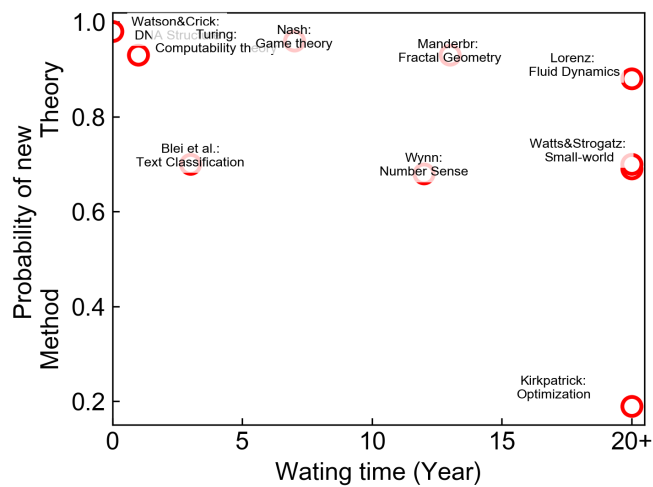

**Figure S7. The shorter waiting time for theory innovation.** The ten most significant breakthrough papers from the survey are analyzed here. The waiting time for theory innovation tends to be shorter than method innovation, which is consistent with the overall results.

We also tried several alternative prompting strategies for the coding, for robustness tests. These strategies include:
  A. Provide a few-shot prompt of concrete examples that guide the model's understanding, instead of the zero-shot prompt. We provide the case of Turing and Church for the example of theory innovation, and the "small world" case for the example of method innovation, using the following prompt, ahead of the original prompt. The results are consistent with that shown in the main text. Adding the few-shots doesn't change the downward slope.

  *Example of "method innovation": [[Paper B, "The small world problem" by Milgram (1967), is a seminal work that introduced the concept of the "small world problem," which suggests that any two people in the world are connected through a short chain of acquaintances. The paper presented empirical evidence for this phenomenon through a series of experiments.\n\nPaper A, "Collective dynamics of \'small-world\' networks" by Watts and Strogatz (1998), builds upon Milgram\'s work by providing a mathematical framework to understand the small-world phenomenon. Watts and Strogatz introduced a new type of network model, known as the "small-world network," which exhibits both local clustering and long-range connections. They also developed a set of mathematical tools to analyze the properties of these networks.\n\nThe key innovation in Paper A is methodological refinement of the small-world concept using graph theory and network analysis. Watts and Strogatz provided a rigorous mathematical framework to study the small-world phenomenon, which was previously described only empirically by Milgram. This formalization enabled the development of new models and simulations to understand the behavior of complex networks.\n\nTherefore, Paper A is an innovation in \'method\']];\nExample of "theory innovation": [[Paper B, "A note on the Entscheidungsproblem" by Church focuses on the Entscheidungsproblem (decision problem) and uses a lambda calculus-based approach to show that there is no general method for determining the validity of a mathematical statement.\n\nIn contrast, Paper A "On computable numbers, with an application to the Entscheidungs problem" by Turing introduces the concept of a machine that can perform computations.\n\nThe key innovation in Paper A is a new theory of computability, which is a quite different conceptual approach from Church\'s approach of understanding the Entscheidungsproblem.\nTherefore, Paper A is an innovation in \'theory\']]*

  B. Providing a prompt with more concrete wording. Instead of asking for theory/method innovation, we also tried a more concrete wording of conceptual difference/formalism difference. The results are consistent.

  *For the following two papers: Paper A {title of paper}, whose abstract is {paper abstract}, and its top reference Paper B {title of reference}, whose abstract is {reference abstract}, Paper A is better considered as (1): an innovation in 'conceptual difference' (a different conceptual angle from its reference paper above), (2): an innovation in 'formalism difference' (an improvement on or formalization of the latter, suh as in methodological or in mathematical frameworks)? Only give the option number.*

  C. Add a third option of "others". We also tried to let the model choose from (1) theory innovation (2) method innovation (3) others, to test for robustness. The results are also consistent.